\documentclass[11pt,twoside]{article}
\usepackage{asp2014}

\aspSuppressVolSlug
\resetcounters

\begin{document}

\title{Full LST-1 data reconstruction with the use of convolutional neural networks}

\author{Jakub~Jury\v{s}ek$^1$, Etienne Lyard$^1$ and Roland Walter$^1$ for the CTA LST project}
\affil{$^1$University of Geneva, Astronomy Department, Geneva, Switzerland; \email{jakub.jurysek@unige.ch}}

\paperauthor{Jakub~Jury\v{s}ek}{jakub.jurysek@unige.ch}{ORCID}{University of Geneva}{Astronomy Department}{Geneva}{}{1200}{Switzerland}
\paperauthor{Etienne Lyard}{jakub.jurysek@unige.ch}{ORCID}{University of Geneva}{Astronomy Department}{Geneva}{}{1200}{Switzerland}
\paperauthor{Roland Walter}{jakub.jurysek@unige.ch}{ORCID}{University of Geneva}{Astronomy Department}{Geneva}{}{1200}{Switzerland}



  
\begin{abstract}
The Cherenkov Telescope Array (CTA) will be the world's largest and most sensitive ground-based gamma-ray observatory in the energy range from a few tens of GeV to tens of TeV. The LST-1 prototype, currently in its commissioning phase, is the first of the four largest CTA telescopes, that will be built in the northern site of CTA in La Palma, Canary Islands, Spain. In this contribution, we present a full-image reconstruction method using a modified \texttt{InceptionV3} deep convolutional neural network applied on non-parametrized shower images. We evaluate the performance of optimized networks on Monte Carlo simulations of LST-1 shower images, and compare the results with the performance of the standard reconstruction method. We also show how both methods work on real-data reconstruction.
\end{abstract}

\section{Introduction}
When primary gamma-ray photons or cosmic-ray particles enter the atmosphere, they interact with atomic nuclei and produce showers of secondary particles emitting Cheren\-kov radiation. In ground-based gamma-ray astronomy, this radiation can be observed by Imaging Atmospheric Cherenkov Telescopes (IACTs), forming roughly elliptical images of the shower in the camera plane of a focusing telescope. In typical IACT observations, the trigger rate from diffuse cosmic-ray background is about 1000x higher than the trigger rate from gamma-ray photons even for the brightest gamma-ray sources in the sky, and therefore a strong background suppression is necessary. Besides reducing this background, the goal of the image analysis is to reconstruct the energy and incoming direction of the primary gamma-ray. In this study, we compare the reconstruction performance of standard Random Forests (RFs) \citep{Breiman2001} with a novel method using a deep convolutional neural network (CNN) on Monte Carlo (MC) simulations \citep{1998cmcc.book.....H, 2008APh....30..149B} and first data from the LST-1 telescope prototype for CTA \citep{doi:10.1142/10986}.

\section{Methods}
We used a modified \texttt{InceptionV3} CNN architecture (Fig.~\ref{inception_architecture}) \citep{Szegedy_2016_CVPR, 2020JPhCS1525a2084L} consisting of a variable number of inception modules followed by five sequential convolutional layers, whose capability of gamma/hadron separation was previously proven by \citet{2020JPhCS1525a2084L}. In this contribution, we study the performance of this network for full event reconstruction. Our networks were trained on MC gamma and MC proton shower images from a standard CTA MC production data set (Prod5) for LST-1. We used \texttt{adadelta} and \texttt{rmsprop} optimizers for gamma/hadron separation and energy/direction reconstruction, respectively. Unlike RFs, which need image parameters on input leading to a loss of information, we trained CNN directly on shower images and times of maxima of waveform in each pixel which follow a temporal gradient, calibrated in \texttt{lstchain v0.6.3} \footnote{\url{https://github.com/cta-observatory/cta-lstchain/releases/tag/v0.6.3}}. In order to get a standard rectangular input image for the \texttt{Keras} Python library, hexagonal images from the telescope had to be reshaped following the method introduced in \citet{2020JPhCS1525a2084L}. Our results suggest that such reshaping doesn't introduce any position reconstruction bias if more than 3 inception modules are used. We also applied relatively weak selection cuts on MC events (Intensity\footnote{The total integrated signal in pixels containing Cherenkov light emitted by the shower particles, after removal of the background pixels.} > 50 p.e., leakage\footnote{The fraction of the shower signal contained in the two outermost pixel rings of the camera.} < 0.2), to create a MC sample that was used for training/testing of CNNs and RFs consistently. The number of inception modules and dropout were optimised for each reconstruction task on a subset of $10^5$ training events (left Fig.~\ref{angular_resolution}). Parameters providing the best results in terms of Area Under the Curve (AUC)\footnote{Area under the ROC curve, usually being used as a measure of the ability of a classifier to distinguish between classes.}, energy and angular resolution were used for final training of each CNN on $5\times10^5$ events.
\articlefigure[width=.99\textwidth]{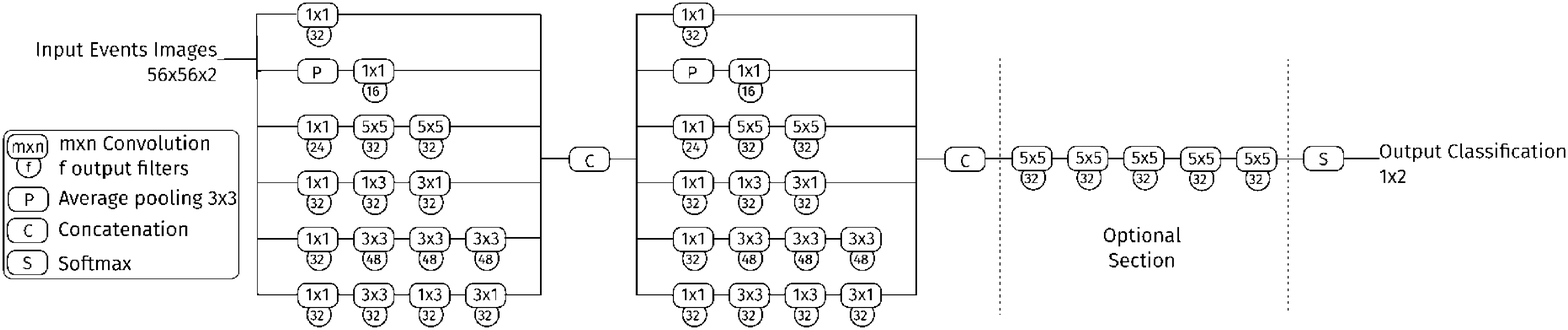}{inception_architecture}{Architecture of modified \texttt{InceptionV3} used in this study. Number of inception modules was optimized for each reconstruction task. Figure from \citet{2020JPhCS1525a2084L}.}

\articlefiguretwo{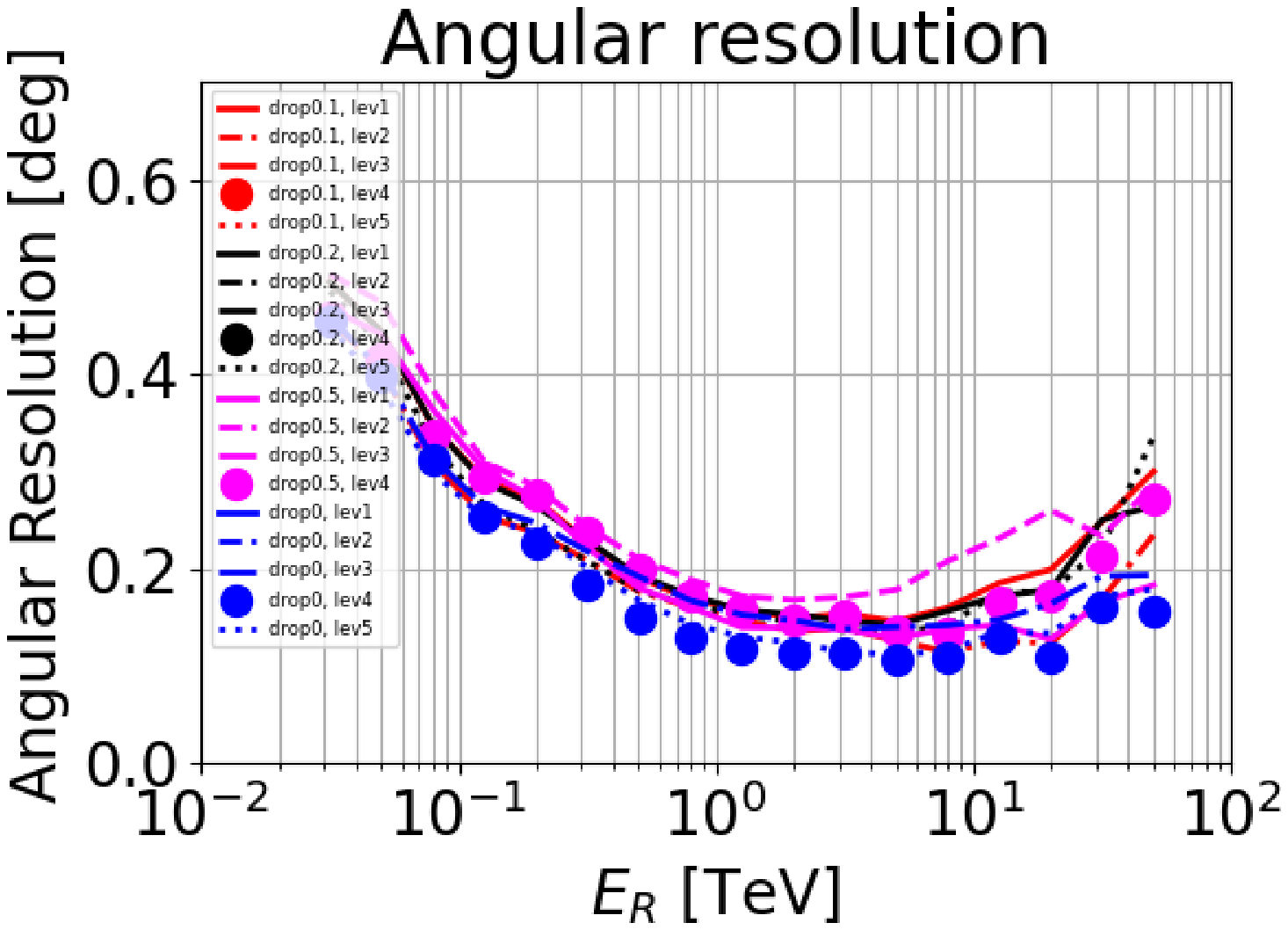}{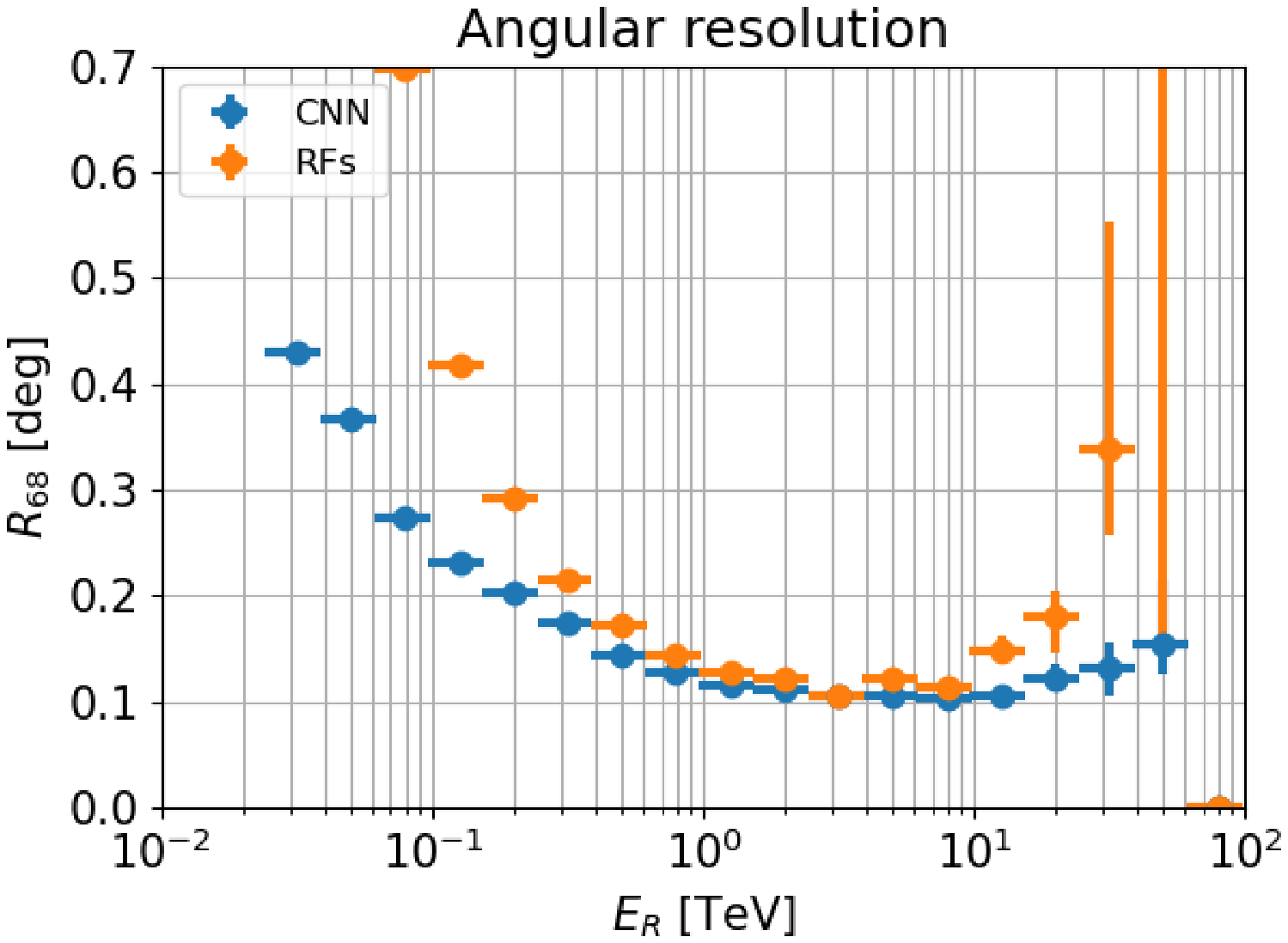}{angular_resolution}{\emph{Left:} An example of angular resolution evaluated on testing point gammas for different parameters of CNN. \emph{Right:} Angular resolution of reconstruction by optimized CNN compared with Random Forests.}

\section{Reconstruction performance}
Figure~\ref{roc_energy} (left) shows receiver operating characteristic curve (ROC) for gamma/hadron separation trained on simulated diffuse gammas and diffuse protons. A comparison with the separation power of RFs trained/tested on Hillas parameters extracted from the same MC sample is also shown, and it can be clearly seen that CNN significantly outperform RFs. Angular and energy resolution curves are shown in Figure~\ref{angular_resolution} (right) and Figure~\ref{roc_energy} (right), respectively. For the mid-energy range the angular resolution improvement gained from CNN is not significant, but it’s very prominent for low and high energies, which can be a hint that CNN are particularly good in the reconstruction of noise-dominated low energy shower images, or showers cut by the camera edges. In energy reconstruction, RFs outperform CNNs by 5\%. The worse reconstruction power of CNN in this case could be caused by image normalisation in the preprocessing step and needs to be addressed in a following study.
\articlefiguretwo{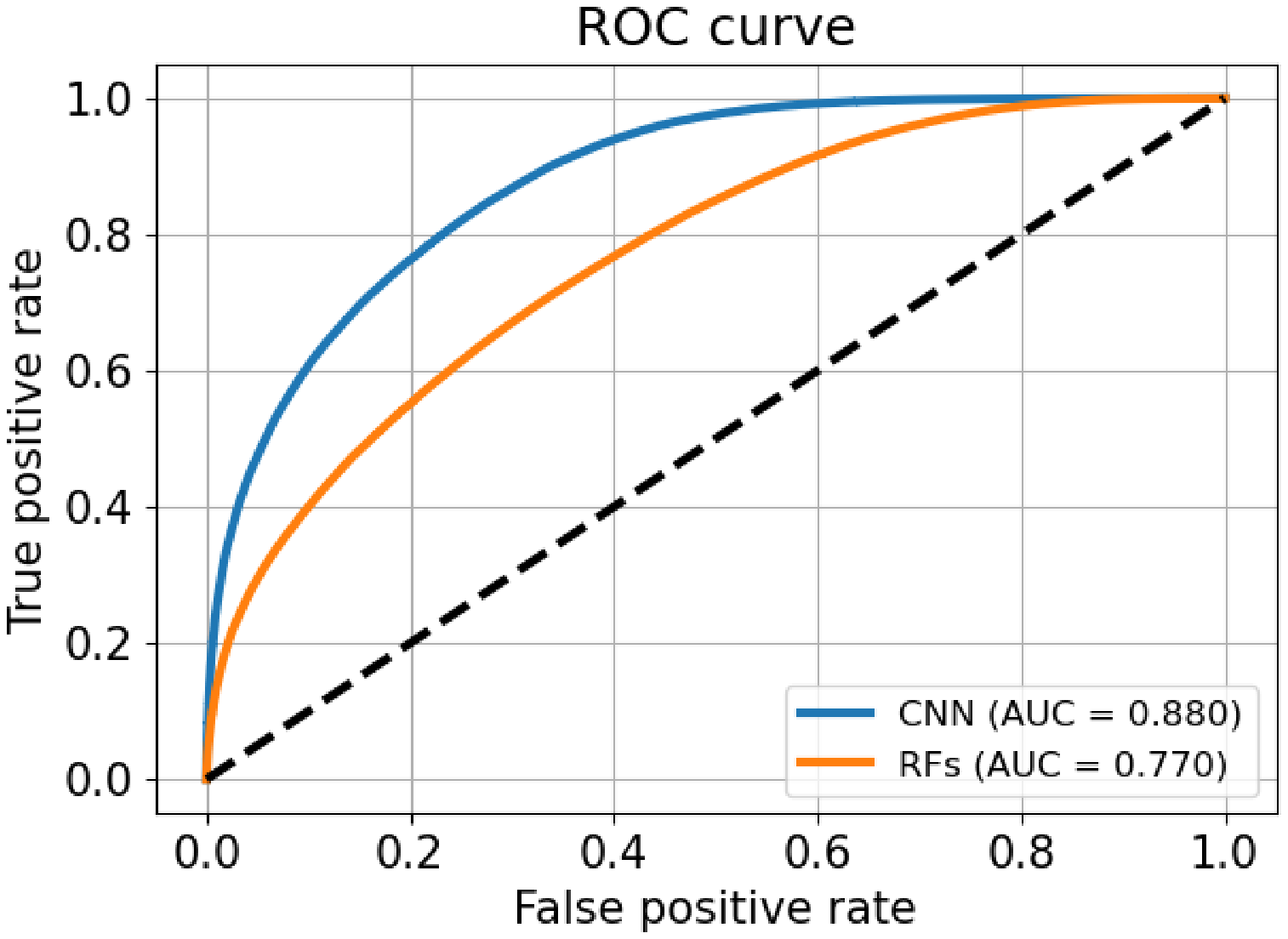}{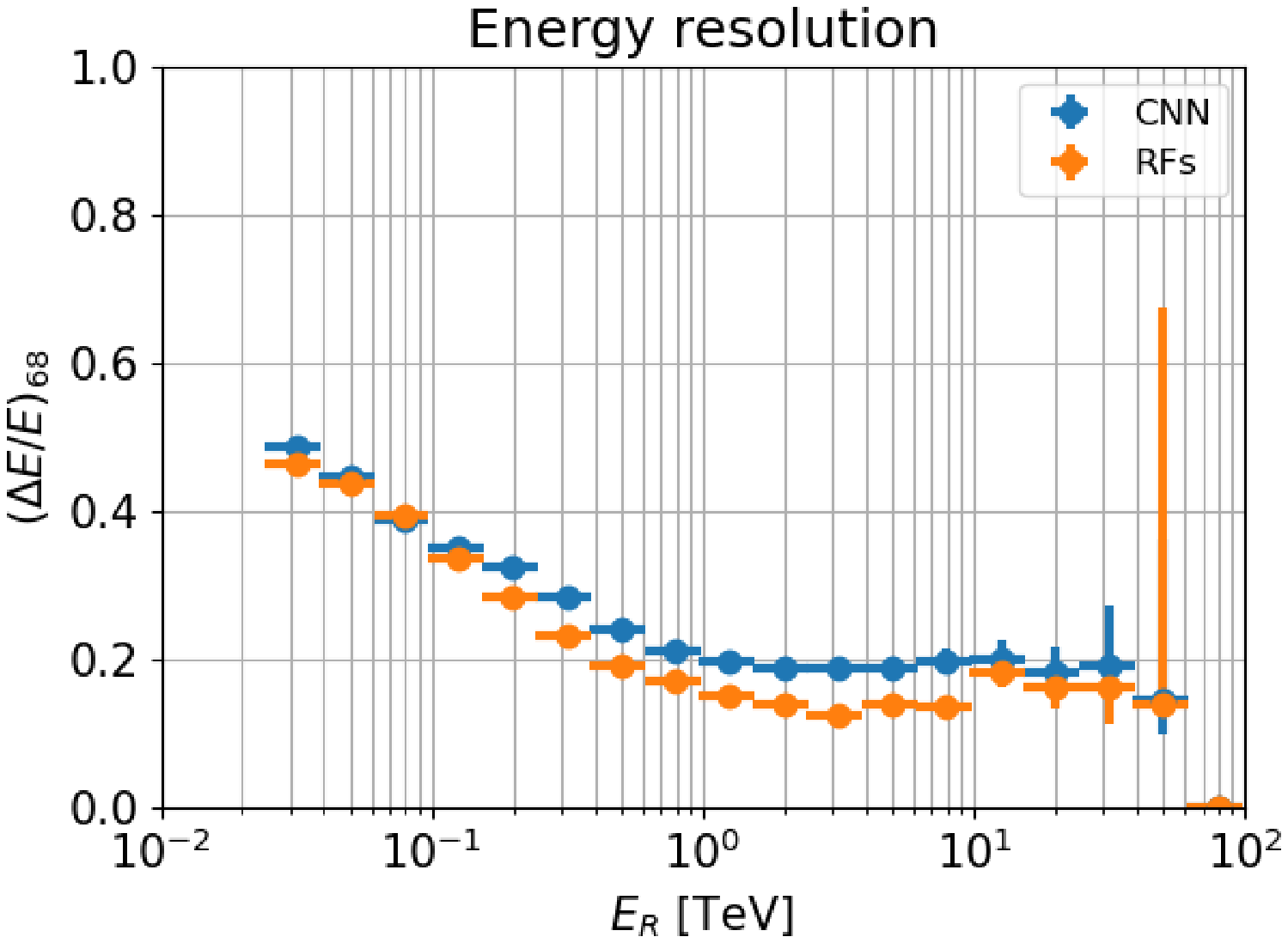}{roc_energy}{\emph{Left:} ROC for gamma/hadron separation evaluated on on-axis point gammas and diffuse protons. \emph{Right:} Energy resolution of CNN reconstruction compared with Random Forests.}

\section{Analysis of real data}
Despite its good performance on simulations, CNN often fails when applied to real data due to subtle differences between the two (e.g. different night sky background level, or stars present in the real field of view), and thus the evaluation of the performance on the real data is essential. We used CNN and RF trained on the same MC events to reconstruct events from Crab Nebula observations from November 2020 (19 mins on-source and 15 mins off-source), and reached a similar detection significance of 6.2$\sigma$ for both methods after carefully matching background levels using an adaptive gammaness cut on the CNN reconstructed data (Gammaness cut for RFs processed data was 0.8), and common $\Theta^2$ < 0.05 $\mathrm{deg}^2$ cut on the source region. A significance map for the CNN reconstructed data is shown in Figure~\ref{hist_sigma} (left). The $\Theta^2$ distribution for both methods (right Fig.~\ref{hist_sigma}) shows that contrary to the performance evaluated on MC, the angular resolution of the CNN on data is worse than we expected, which is probably caused by MC-data differences, which will be addressed in a following study.
\articlefiguretwo{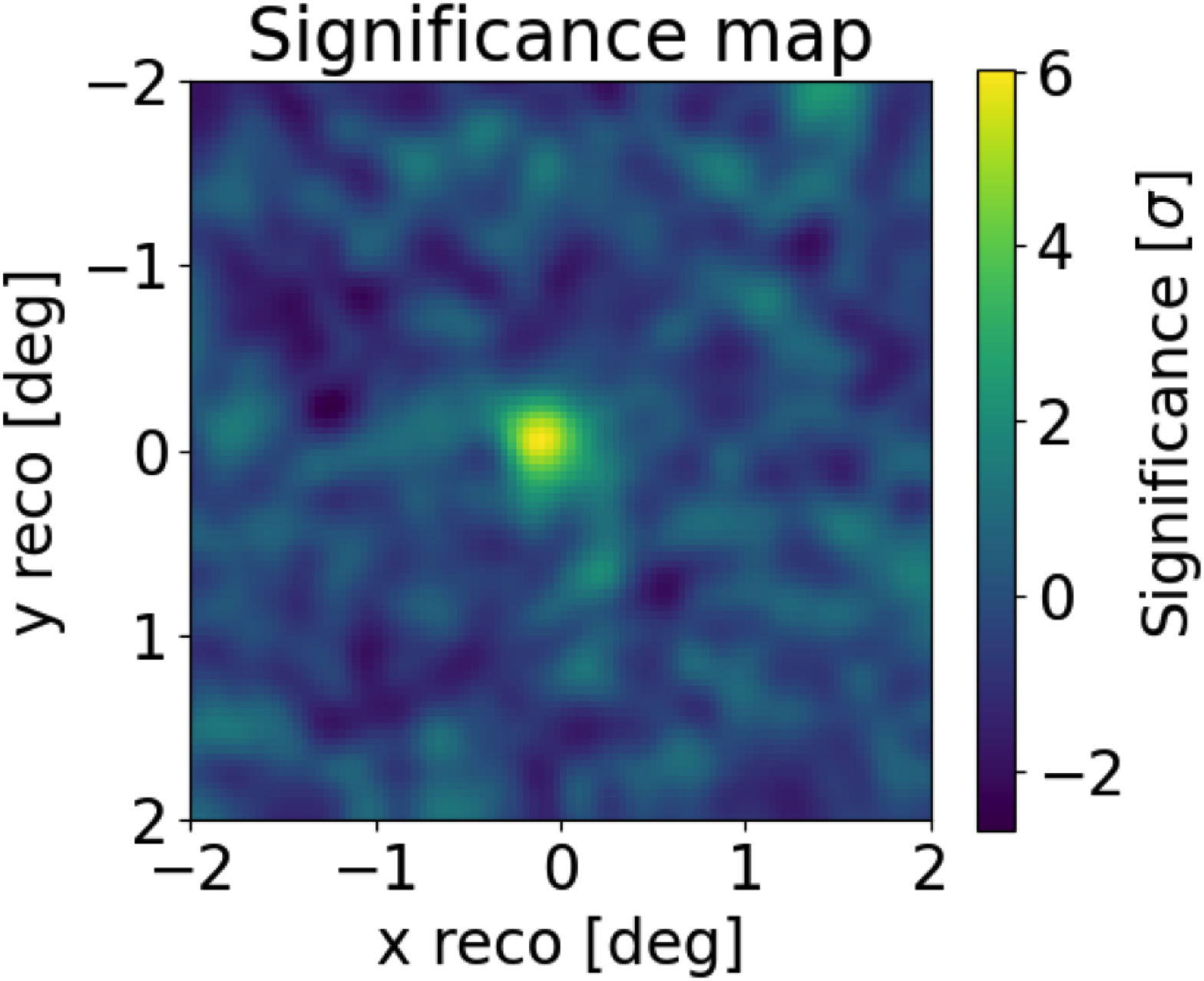}{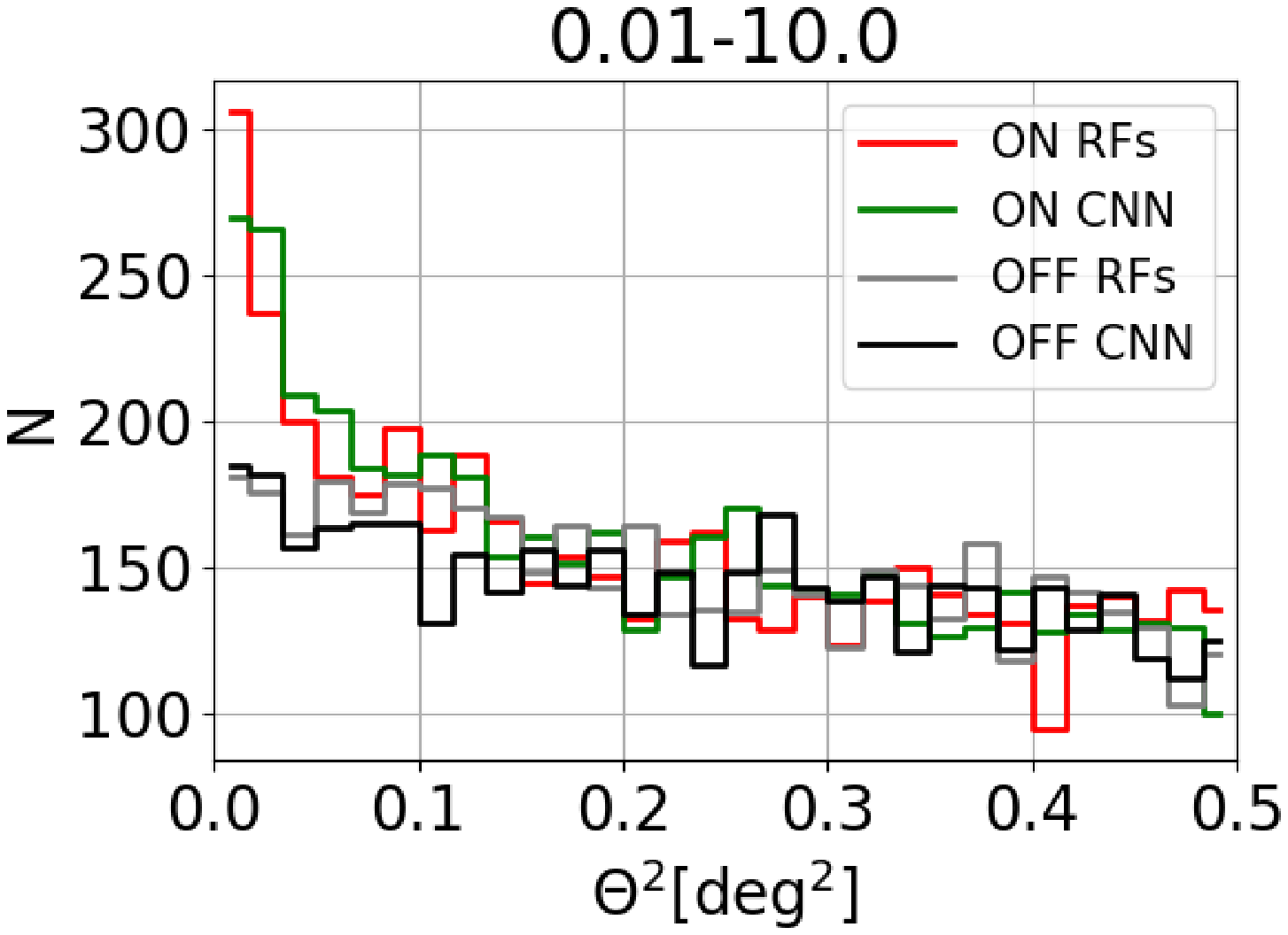}{hist_sigma}{\emph{Left:} Significance map of events reconstructed with CNN. \emph{Right:} $\Theta^2$ distribution of events in energy range 0.01- 10 TeV.}

\section{Conclusions}
In this contribution, we demonstrated a potential of CNNs to be used not only for gamma/hadron separation, but also for full event reconstruction with a competitive performance to the standard RFs reconstruction. Even though the CNN performance on real data seems to be slightly worse than expected from the full MC study, there is much room for improvement, particularly in MC-data tuning.

\acknowledgements We gratefully acknowledge financial support from the agencies and organizations listed here: www.cta-observatory.org/consortium\_acknowled\-gments. We would like also to thank CSCS in Lugano for the computer hours on PizDaint computer.

\bibliography{X1-001}


\end{document}